\documentclass[aps,floats,twocolumn]{revtex4}
\usepackage{tabularx,graphicx}
\usepackage{epsfig}
\usepackage{amsmath}
\usepackage{bbm}
\usepackage{graphicx}

\begin{document}
%\draft
%\flushbottom
%\twocolumn[
%\hsize\textwidth\columnwidth\hsize\csname @twocolumnfalse\endcsname

\title{Unconventional quasiparticle lifetime in undoped graphene}
\author{J. Gonz\'{a}lez$^a$ and E. Perfetto$^b$ \\}
\address{
        $^a$Instituto de Estructura de la Materia,
        Consejo Superior de Investigaciones Cient\'{\i}ficas, Serrano 123,
        28006 Madrid, Spain\\
        $^b$Dipartimento di Scienza dei Materiali, Universit\`a di Milano-Bicocca, 
        Via Cozzi 53, 20125 Milano, Italy}

\date{\today}

\begin{abstract}
%\widetext
We address the question of how small can the quasiparticle decay rate be at low
energies in undoped graphene, where kinematical constraints are known to prevent 
the decay into particle-hole excitations. For this purpose, we study the 
renormalization of the phonon dispersion by many-body effects, which turns out 
to be very strong in the case of the out-of-plane phonons at the $K$ point of the 
spectrum. We show that these evolve into a branch of very soft modes that provide 
the relevant channel for quasiparticle decay, at energies below the scale of the 
optical phonon modes. In this regime, we find that the decay
rate is proportional to the cube of the quasiparticle energy. This implies that
a crossover should be observed in transport properties from the linear dependence
characteristic of the high-energy regime to the much slower decay rate due to the
soft phonon modes.

\end{abstract}
%\pacs{71.10.Pm,73.22.-f}

%]
\maketitle

%\narrowtext
%\tightenlines

%\newpage

The recent fabrication of single atomic layers of carbon has attracted 
a lot of attention, as this material (so-called graphene) provides
the experimental realization of a system where the low-energy electronic
excitations behave as massless Dirac fermions\cite{geim,kim}. 
The main results reported in
Refs. \onlinecite{geim} and \onlinecite{kim} can be understood as a 
reflection of the linear dependence on momentum of the quasiparticle energy
$\varepsilon ({\bf k})$. The honeycomb lattice structure of graphene is 
actually known to lead to a bandstructure with conical shape 
$\varepsilon ({\bf k}) = \pm v_F |{\bf k}|$ around the corners of the 
hexagonal Brillouin zone, with conduction and valence bands meeting at
the Dirac points.
The relativistic-like invariance arising from the massless Dirac
quasiparticles has been shown to be at the origin of a number of 
remarkable electronic properties, like the finite lower bound of the 
conductivity at the charge neutrality point\cite{paco,kat,twor,mac}, 
the anomalous integer Hall effect\cite{paco,ando,gus}, 
and the absence of backscattering in the presence of long-range 
scatterers\cite{suzu}.

More recently, the many-body properties of the graphene layer have been
also investigated. One of the relevant issues addressed is whether the 
quasiparticle properties have to correspond to the expected behavior 
for a Fermi liquid in two dimensions\cite{prbr,sarma}. In this 
regard, it has been pointed out that the $e$-$e$ interactions lead 
to quite different quasiparticle features in graphene depending on whether 
the material is doped or not\cite{sarma}. This can be understood from the 
particular kinematical constraints of the conical dispersion, that prevent 
the Dirac quasiparticles from decaying into interband particle-hole 
excitations\cite{hwang}.  

In doped graphene, intraband processes are responsible for the 
quasiparticle decay, leading to a quadratic dependence on energy of the 
decay rate\cite{sarma}. 
On the other hand, when the Fermi level is at the charge 
neutrality point, the electron self-energy has a linear dependence on 
frequency. However, the maximum energy released in the scattering of a 
quasiparticle with momentum transfer ${\bf q}$ is at the boundary of the 
continuum of particle-hole excitations, which have energy $\ge v_F |{\bf q}|$. 
In situations where the Coulomb interaction remains singular in the limit 
${\bf q} \rightarrow 0$, as it happens in the layers of bulk graphite, 
a finite spread in the momentum of the quasiparticles is enough to give 
rise to a finite decay rate\cite{unconv}. 
This mechanism must rely however on some 
effect extrinsic to the 2D system (disorder, for instance), and it would be 
absent anyhow as soon as the Coulomb interaction is screened beyond a 
certain distance.

In this paper we address the question of whether the quasiparticle decay 
rate may actually vanish in a graphene layer with the
Fermi level tuned at the charge neutrality point. This study is relevant 
as it faces the possibility of having an electron liquid made of extremely 
long-lived quasiparticles. Thus, we will look for many-body effects which 
may give rise to suitable gapless excitations and consequent quasiparticle 
decay channels in undoped graphene.

It is known that, in the absence of doping, the 2D system does not support 
plasmon excitations\cite{unconv}. 
Yet the polarization of the electron liquid is singular
at low energies, and this may be the source of potential instabilities.
We will see that there is actually a significant renormalization 
of the interactions at the large momentum-transfer $K$ connecting
the two inequivalent Dirac points in graphene. At such large momentum, the 
singular behavior of the electron polarization tends to amplify the effects 
of the electron-phonon interaction, which prevails over the 
Coulomb interaction. Thus, we will show that gapless phonon branches may 
appear at the $K$ point when graphene is lying on a substrate. The resulting
low-energy phonon modes provide then the relevant mechanism for the decay of 
quasiparticles in undoped graphene, though with a very low decay
rate that turns out to be proportional to the cube of the quasiparticle 
energy.

We begin by considering the hamiltonian for Dirac quasiparticles in
graphene, at energies below the scale of $\sim 1$ eV for which the 
dispersion can be taken as linear:
\begin{equation}
H_0 =   v_F \int d^2 k \; \Psi^{(a) \dagger} ({\bf k})
   \: \mbox{\boldmath $\gamma $}^{(a)} \cdot {\bf k} \:
              \Psi^{(a)} ({\bf k})
\end{equation}
In the above expression, a sum is implicit over the index $a$ accounting 
for the two different valleys and corresponding Dirac spinors $\Psi^{(a)} $ 
at opposite corners $K, -K$ in the graphene Brillouin zone.
$\mbox{\boldmath $\gamma $}^{(a)}$ are different sets of Pauli matrices for 
$a = 1, 2$, which must be chosen according to the appropriate chirality 
of the modes at $K, -K$ as 
$\mbox{\boldmath $\gamma $}^{(1)} \equiv ( \sigma_x , \sigma_y )$, 
$\mbox{\boldmath $\gamma $}^{(2)} \equiv ( -\sigma_x , \sigma_y )$ \cite{ando}.
As a first step in the development of the many-body theory, we will
assume that the quasiparticles interact through a Coulomb potential 
\begin{equation}
V_0 ({\bf q}) = \frac{e^2}{ 2\kappa |{\bf q}| }
\label{pot}
\end{equation}
with a dielectric constant $\kappa $ dictated by the coupling to the substrate.

As is well-known, the quasiparticles of the 2D system provide very limited 
screening of the Coulomb potential in (\ref{pot}). This effect
can be assessed by computing the polarization 
\begin{eqnarray}
 \Pi_0^{(a,b)} ({\bf q}, i \overline{\omega}_q )   & = &   
  4 \: {\rm Tr }    \int \frac{d^2 k}{(2 \pi)^2}    
   \int \frac{d \overline{\omega}_k}{2 \pi}  \:              \nonumber      \\
  &  &  G^{(a)} ({\bf k}+{\bf q}, i\overline{\omega}_k + i\overline{\omega}_q)  
     \: G^{(b)} ({\bf k}, i\overline{\omega}_k )          
\label{pol}
\end{eqnarray}
with Dirac propagators $G^{(a)} ({\bf k}, i \overline{\omega}_k ) = 
 1/( i \overline{\omega}_k - 
  v_F \mbox{\boldmath $\gamma $}^{(a)} \cdot {\bf k} )$.
At small momentum-transfer, the trace in (\ref{pol}) is taken over excitations
in the same valley $a=b$, with the result that
${\rm Tr } ( i \overline{\omega}_q + 
  v_F \mbox{\boldmath $\gamma $}^{(a)} \cdot {\bf q} )
  ( i \overline{\omega}_k + 
  v_F \mbox{\boldmath $\gamma $}^{(a)} \cdot {\bf k} )  = 
  - 2\overline{\omega}_q \overline{\omega}_k  + 2 v_F^2 {\bf q} \cdot {\bf k} $.
This leads to an expression for the intravalley polarization
$ \Pi_0^{(a,a)} ({\bf q}, i \overline{\omega}_q ) =  - {\bf q}^2 / 
 8 \sqrt{v_F^2 {\bf q}^2 + \overline{\omega}_q^2 }$ \cite{np}. Going back to  
real frequency $\omega_q = i \overline{\omega}_q $, we find a divergence
of the polarization at $\omega_q = v_F |{\bf q}|$. This marks actually the
threshold for the creation of particle-hole pairs in the electron liquid. 
The particle-hole continuum is above the maximum energy 
$v_F |{\bf q}|$ released in the scattering of a quasiparticle with 
momentum transfer ${\bf q}$. This explains that the quasiparticle decay into 
particle-hole pairs is forbidden in the case of undoped graphene. 

In the case of intervalley scattering of quasiparticles, the polarization is
also affected by a similar divergence at $\omega_q = v_F |{\bf q}|$, where
${\bf q}$ stands now for a small deviation around the large momentum $K$. The
computation of the polarization (\ref{pol}) with $a \neq b$ leads to the
trace  ${\rm Tr } ( i \overline{\omega}_q + 
  v_F \mbox{\boldmath $\gamma $}^{(a)} \cdot {\bf q} )
  ( i \overline{\omega}_k + 
  v_F \mbox{\boldmath $\gamma $}^{(b)} \cdot {\bf k} )  = 
- 2\overline{\omega}_q \overline{\omega}_k  - 2 v_F^2 q_x k_x + 2 v_F^2 q_y k_y $.
This can be assimilated to the above computation for $a = b$
if the $y$ component of each momentum is exchanged with the frequency 
$\overline{\omega }$, and an overall $-$ sign is introduced. It can be checked
by direct calculation that the result for the intervalley polarization 
corresponds actually to operating that transformation in the above expression 
for $ \Pi_0^{(a,a)}$, that is,
$ \Pi_0^{(1,2)} ({\bf q}, i \overline{\omega}_q ) = 
  (p_x^2 + \overline{\omega}_q^2 / v_F^2) /
     8 \sqrt{v_F^2 {\bf q}^2 + \overline{\omega}_q^2 }  $.
The polarization thus obtained shows a preferred direction in momentum space,
which is a reflection of having considered the scattering between
two Dirac valleys along the $x$ direction. The result physically sensible can
be obtained by averaging over the processes involving the three equivalent 
nearest-neighbor valleys of the $K$ point. These include in particular the 
valleys rotated by an angle of $\pm 2\pi /3 $ with respect to the $x$-axis.
Taking into account the three different contributions, we 
get the final result for the intervalley polarization 
\begin{equation}
\widetilde{\Pi}_0 ({\bf q}, \omega_q ) = 
  \frac{{\bf q}^2 /2 - \omega_q^2 / v_F^2}
      {8 \sqrt{v_F^2 {\bf q}^2 - \omega_q^2 } }
\label{inter}
\end{equation}

The polarization (\ref{inter}) does not give rise to any singularity 
when renormalizing the Coulomb interaction, as the Coulomb potential gets 
dressed at large momentum-transfer $K$ in the form 
$V_0 ({\bf q}) \approx e^2/(2\kappa K - \widetilde{\Pi}_0 ({\bf q}, \omega_q ))$. 
On the contrary, the singular behavior of the intervalley polarization may 
lead to important effects in the phonon sector. This consideration is relevant 
for lattice vibrations coupling to the total electron charge, as it happens 
in the case of the out-of-plane phonons. The electron-phonon interaction can 
be analyzed in terms of the atomic deformation potential induced by the 
lattice vibrations\cite{adp}. When graphene is lying on a substrate,
the mirror symmetry of the vibrations perpendicular to the carbon layer is
broken, and the on-site deformation potential induces a linear coupling of
the out-of-plane phonons to the total electron charge. If we denote by 
$c^{\dagger}_{\bf q}, c_{\bf q}$ the creation and annihilation operators 
for the modes around the $K$ point of a branch of out-of-plane 
phonons, we can describe the phonon sector by means of
kinetic and interaction terms in the hamiltonian:
\begin{eqnarray}
H_{\rm ph}  & = &  \int d^2 q 
 \: \omega_0 ({\bf q}) \:   c^{\dagger}_{\bf q} c_{\bf q}    \nonumber  \\ 
H_{\rm e-ph}  & = &   g \int d^2 k d^2 q
 \Psi^{(a) \dagger}({\bf k+q}) \Psi^{(b)} ({\bf k}) 
   ( c_{\bf q} + c^{\dagger}_{-{\bf q}}  )      \;\;\;\;     
\label{ep}
\end{eqnarray}
For the characterization of the phonon branch, it will be enough to 
approximate the energy of the out-of-plane phonons about the $K$ point
by $\omega_0 \approx 70$ meV. 
The electron-phonon coupling $g$ can be obtained as the atomic deformation
potential (of the order of a few eV) times $1/\sqrt{m_C \omega_0}$
($m_C$ being the carbon atomic mass) \cite{adp}.

\begin{figure}
\begin{center}
\mbox{\epsfxsize 6.5cm \epsfbox{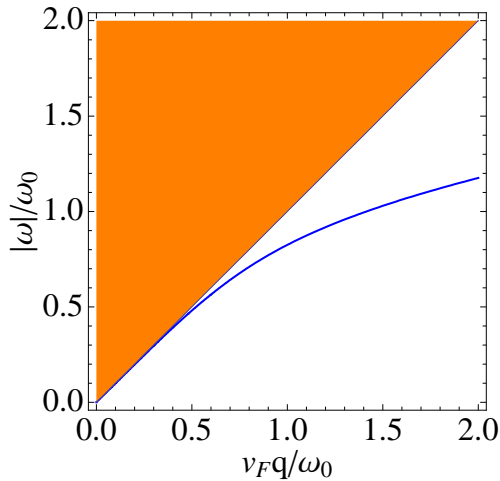}}
\end{center}
\caption{Plot of the region corresponding to the particle-hole continuum 
(shaded area) and the phonon dispersion arising from the solution of Eq. 
(\ref{pole}).}
\label{one}
\end{figure}

The intervalley polarization induces a strong renormalization of the 
out-of-plane phonons in graphene. It has been already pointed out that the
interaction with the electronic degrees of freedom may give rise to significant
Kohn anomalies in the dispersion of in-plane optical phonons\cite{ka1,ka2}. 
The coupling to
these branches is given in general by the modulation of the transfer integral
between nearest-neighbor atoms in the carbon lattice. In our framework, this 
gives rise to an electron-phonon vertex proportional to the matrix $\sigma_x $. 
When introduced in the computation of the trace in the polarization, such a 
vertex gives rise to simple scalar products in $({\bf q}, \overline{\omega}_q )$ 
space, leading to a susceptibility  proportional to 
$\sqrt{v_F^2{\bf q}^2 - \omega_q^2 }$. However, in the case of phonons  
coupling to the total electron charge, the particle-hole polarization (\ref{inter}) 
induces a more profound anomaly in the phonon dispersion. If we approximate 
the bare phonon propagator about the $K$ point by $D_0 ({\bf q}, \omega ) \approx 
2 \omega_0 /(\omega^2 - \omega_0^2 + i \epsilon )$, the renormalized 
propagator $D ({\bf q}, \omega )$ dressed with the particle-hole polarization
becomes
\begin{equation}
D ({\bf q}, \omega ) \approx \frac{2 \omega_0 }{\omega^2 - \omega_0^2 + 
    i \epsilon  -  2 \omega_0 g^2 \widetilde{\Pi}_0 ({\bf q}, \omega  ) }
\label{dr}
\end{equation}
The renormalized phonon energies are found by setting to zero the denominator 
of the propagator (\ref{dr}), which leads to the equation
\begin{equation}
\omega^2 - \omega_0^2 
    - \frac{g^2}{ v_F^2}  \omega_0 \frac{v_F^2{\bf q}^2 /2 - \omega^2 }
          {4 \sqrt{v_F^2 {\bf q}^2 - \omega^2}}  =  0
\label{pole}
\end{equation}
It can be checked that the phonon dispersion thus computed becomes 
gapless at ${\bf q} = 0$, adopting the form of a low-energy branch
below the continuum of particle-hole excitations as shown in Fig. \ref{one}. 
At this point, it becomes pertinent however to 
assess the effects of the Coulomb interaction on the renormalization of the
phonon properties.

The Coulomb and the phonon-mediated interaction have to be considered on the 
same footing when analyzing their role in the renormalization of the phonon
propagator. Thus, we can define an intervalley particle-hole susceptibility
$\widetilde{\Pi} ({\bf q}, \omega_q )$ dressed by the effect of the Coulomb 
interaction and, therefore, satisfying the equation
\begin{equation}
\widetilde{\Pi} = \widetilde{\Pi}_0 + \widetilde{\Pi}_0 V_0  \widetilde{\Pi} 
\label{susc}
\end{equation}
In terms of this susceptibility, the renormalized phonon propagator 
$D ({\bf q}, \omega )$ can be obtained as 
\begin{equation}
D^{-1} = D_0^{-1} - g^2  \widetilde{\Pi}
\label{full}
\end{equation}
Combining the solution of (\ref{susc}) with (\ref{full}), we arrive at the final
expression
\begin{equation}
D ({\bf q}, \omega ) = \frac{2 \omega_0 (1 - \frac{e^2}{2\kappa K} 
                                  \widetilde{\Pi}_0 ({\bf q}, \omega ))}
{\omega^2 - \omega^2_0 +  i \epsilon   - ((\omega^2 - \omega^2_0)\frac{e^2}{2\kappa K} 
                  + 2\omega_0 g^2) \widetilde{\Pi}_0 ({\bf q}, \omega )}
\label{prop}
\end{equation}
The strength of the phonon-mediated interaction is given by the dimensionless 
coupling $g^2/v_F^2$, which can be estimated as $\sim 0.1$. This is smaller 
than the strength of the Coulomb interaction in typical graphene samples, 
where $e^2/\kappa v_F \sim 1$. However, the latter enters in the denominator 
of (\ref{prop}) with a relative weight $\omega_0 /4v_F K $, which is of the 
order of $\sim 0.001$. Therefore, we see that the Coulomb interaction is not 
able to balance the effect of the strong renormalization of the phonon 
propagator at the large momentum transfer $K$.

We find then that, when graphene is tuned at the charge neutrality point, there 
are phonon modes with very low energy around the $K$ point, arising as a 
consequence of the strong renormalization from the coupling to particle-hole 
excitations. Strictly speaking, the phonon branches become gapless only at 
zero temperature in the undoped system. Away from the charge neutrality point
and at finite temperature, the divergence of the polarization in (\ref{inter})
will be cut off by either the thermal energy or the effective chemical
potential of the system. Above such infrared scales, the phonon dispersion will
follow anyhow the trend represented in Fig. \ref{one}. The phonon energy 
$\omega_{\rm ph} ({\bf q})$ obtained from the renormalized propagator is 
actually given by
\begin{equation}
\omega_{\rm ph}  ({\bf q}) \approx 
     v_F |{\bf q}| - \left( \frac{g^2}{8 v_F^2 } \right)^2 
   \frac{v_F^3 |{\bf q}|^3 }{2 \omega_0^2 }   + \ldots
\label{linear}
\end{equation}
We stress that this phonon branch lies in any event away from the 
continuum of particle-hole excitations, opening the possibility to observe 
well-defined phonon modes of very low energy at the $K$ point of graphene.

The existence of the soft phonon branch leads to a channel for the decay 
of quasiparticles in undoped graphene. The maximum energy that can be released 
by a quasiparticle in a scattering process at any low momentum-transfer  
is enough to hit the phonon branch (\ref{linear}), 
so that electron quasiparticles can decay into this type of phonon modes down 
to arbitrarily low energies (at zero temperature). The quasiparticle decay rate
$\tau^{-1} $ can be computed from the electron self-energy 
$\Sigma^{(a)} ({\bf k}, \omega_k)$ as
\begin{eqnarray}
\lefteqn{\tau^{-1} = - {\rm Im} \; \Sigma^{(a)} ({\bf k}, v_F |{\bf k}| ) }    \nonumber  \\
  &    &  \approx   {\rm Im} \; i g^2 \int \frac{d^2 q}{(2 \pi )^2} 
     \frac{d \omega_q }{2 \pi }
   G^{(b)} ({\bf k}-{\bf q}, v_F |{\bf k}| - \omega_q) 
      D ({\bf q}, \omega_q )  \;\;\;\;\;\;\; 
\label{im}
\end{eqnarray}
which amounts to making the convolution of the imaginary part of the electron 
propagator with that of $D ({\bf q}, \omega_q )$.

In Eq. (\ref{im}), the imaginary part of $G^{(b)}$ enforces the 
constraint $\omega_q = v_F |{\bf k}| - v_F |{\bf k} - {\bf q}|$. For that 
frequency, the phonon propagator picks up an imaginary contribution only from 
the phonon branch (\ref{linear}). We have actually
\begin{equation}
\tau^{-1} \approx \frac{\pi }{2} g^2 \int \frac{d^2 q}{(2 \pi )^2}  
   \; \delta  ( Q({\bf q}, \Omega_{{\bf q}}) )
\end{equation}
where $\Omega_{{\bf q}} \equiv v_F|{\bf k}| - v_F|{\bf k} - {\bf q}|$ and
\begin{equation}
Q({\bf q}, \Omega_{{\bf q}}) = 
     \frac{\Omega_{{\bf q}}^2 - \omega_0^2}{2 \omega_0} 
    - \frac{g^2}{v_F^2}   \frac{ v_F^2{\bf q}^2/2 - \Omega_{{\bf q}}^2 }
         { 8\sqrt{v_F^2 {\bf q}^2 - \Omega_{{\bf q}}^2} }
\end{equation}
The integral over ${\bf q}$ can be done by trading the azimuthal 
variable of integration $\phi $ by $\Omega_{{\bf q}}$. Thus we get
\begin{eqnarray}
\tau^{-1}   &  \approx  & 
      \frac{1}{2 \pi } g^2 \int_0^{|{\bf k}|} dq \: |{\bf q}|
 \int_0^{|{\bf q}|} d \Omega_{{\bf q}}                    \nonumber     \\ 
&   &   \;\;\;\;\;\;\; 
     \left| \frac{\partial \phi }{\partial \Omega_{{\bf q}}} \right| 
  \;  \left| \frac{\partial Q }{\partial \Omega_{{\bf q}}} \right|^{-1}
   \delta (\Omega_{{\bf q}} - \omega_{\rm ph}  ({\bf q}) )   
\label{tau}
\end{eqnarray} 

The expression (\ref{tau}) leads to different behaviors depending on whether
the quasiparticle energy is well above or below the scale $\omega_0$. In the
range where $v_F |{\bf k}| \gg \omega_0$, it is easy to see that the Jacobian
$|\partial \phi  / \partial \Omega_{{\bf q}}|$ scales as 
$\sim  |{\bf k}|/ |{\bf q}| \sqrt{4{\bf k}^2 - {\bf q}^2}$, while 
$|\partial Q / \partial \Omega_{{\bf q}}|^{-1}$ does not scale with momentum.
The quasiparticle decay rate shows then a linear dependence on energy 
$\tau^{-1} \sim (g^2/v_F^2)v_F |{\bf k}|$, in agreement with previous 
analyses of the decay due to optical phonons\cite{giust}. On the other hand, 
when the quasiparticle energy is below $\omega_0$, we find that 
$|\partial \phi  / \partial \Omega_{{\bf q}}|$ scales as 
$\sim \omega_0 \sqrt{|{\bf k}| - |{\bf q}|} / \sqrt{|{\bf k}|} v_F^2{\bf q}^2$. 
Moreover, we also have 
$|\partial Q / \partial \Omega_{{\bf q}}|^{-1} \sim v_F^3|{\bf q}|^3/\omega_0^3$.
We get then a decay rate 
\begin{equation}
\tau^{-1}  \approx  \frac{1}{16 \pi }
  \frac{g^4}{v_F} \frac{|{\bf k}|^3}{\omega_0^2} \int_0^1 dx \: x^2 \sqrt{1-x}
\label{cube}
\end{equation}
We arrive at the result that, in the case of undoped graphene,
the quasiparticle decay rate cannot vanish at low energies, even below the 
scale $\omega_0$ of the out-of-plane phonons, where it must be proportional 
to the cube of the quasiparticle energy. 

The behavior (\ref{cube}) differs significantly from the rate obtained for a 
long-range Coulomb interaction, which is proportional to the quasiparticle 
energy\cite{sarma,unconv}. We remark that the two behaviors correspond actually 
to quite different conditions. When the Coulomb interaction remains 
long-ranged, as in the layers of bulk graphite, the singular character of the 
potential (\ref{pot}) leads to a jump in the electron self-energy at 
$\omega_k = v_F |{\bf k}|$. A spread in momentum of the quasiparticles (induced 
for instance by disorder) may be invoked to obtain a finite quasiparticle decay 
rate by taking the limit $\omega_k = v_F |{\bf k}| + 0^+$. 
We have to bear in mind however that, in graphene, the divergence of the Coulomb 
potential may be cut off by some finite screening length $l$. If we assume a 
potential of the form $V_0 ({\bf q}) = e^2/\sqrt{ {\bf q}^2 + l^{-2} }$, we can 
estimate the quasiparticle decay rate to lowest order in the Coulomb 
interaction as 
\begin{eqnarray}
\lefteqn{\tau^{-1}    \sim   \lim_{\epsilon \rightarrow 0 } 
                e^4  \int_0^{|{\bf k}|} dq \: |{\bf q}|
 \int_{|{\bf q}|}^{|{\bf q}|+ \epsilon} d \Omega_{{\bf q}}  }    \nonumber     \\
 &  &   \frac{\sqrt{|{\bf k}| - |{\bf q}|}}{\sqrt{|{\bf k}| |{\bf q}|} 
                 \sqrt{\epsilon - (\Omega_{{\bf q}} - v_F |{\bf q}|)}}
  \frac{{\bf q}^2}{({\bf q}^2 + l^{-2}) \sqrt{\Omega_{{\bf q}}^2 - v_F^2 {\bf q}^2}}
          \;\;\;\;\;\;\;
\label{c3}
\end{eqnarray}      
We see that the decay rate in this approach turns out to be  
proportional to $(e^2/v_F)^2 v_F l^2 |{\bf k}|^3$. It is only after sending
the screening length to infinity that the divergence of the Coulomb potential
at ${\bf q} = 0$ is able to change the scaling of the integrand in (\ref{c3}), 
leading to a decay rate proportional to the quasiparticle energy.

In conclusion, we have shown that the dispersion of phonons that couple to the 
total electron density undergoes a strong renormalization when graphene is 
tuned to the charge neutrality point. This reflects in the behavior of the 
out-of-plane phonons, which turn out to develop a gapless branch at the $K$ 
point of the spectrum. In situations where graphene is very lightly doped, 
the mentioned renormalization will still give rise to a branch of very soft 
phonons, with a small gap proportional to the effective chemical potential (as 
measured from the charge neutrality point). We have seen that this branch
provides the relevant channel for the decay of quasiparticles below the 
typical scale of the out-of-plane phonons, $\omega_0 \approx 70$ meV. Thus, 
even at such low energies, the quasiparticle decay rate does not vanish in 
undoped graphene, though it becomes very suppressed, with a dependence 
proportional to the cube of the quasiparticle energy. The present analysis
may then be useful to interpret the results of transport experiments in 
undoped or very lightly doped graphene, where a crossover should be observed
from the linear decay rate characteristic of the high-energy regime to the
much slower cubic dependence found in the paper.

\acknowledgments

%{\center \bf Acknowledgements}
 \noindent
We thank F. Guinea and F. Sols for very fruitful discussions.
The financial support of the Ministerio de Educaci\'on y Ciencia
(Spain) through grant FIS2005-05478-C02-02 is gratefully
acknowledged. E.P. is also financially supported by CNISM and
Fondazione Cariplo-n.Prot.0018524.

\end{document}